\def\transmittedSignal[#1]{p\left(#1\right)}

\def\basisFunction[#1]{B_{#1}(\timeVar)}
\def\amountOfShift[#1]{\tau_{#1}}
\def\dataSymbols[#1]{d_{#1}}
\def\angleSignal[#1]{\psi_{#1}(\timeVar)}
\def\instantaneousFrequency[#1]{F_{#1}(t)}
\def\besselFunctionFirstKind[#1][#2]{J_{#1}\left(#2\right)}

\def\fourierSeries[#1]{c_{#1}}

\def\timeVar{t}

\def\fresnelC[#1]{C(#1)}
\def\fresnelS[#1]{S(#1)}

\newcommand\mydots{\hbox to 1em{.\hss.\hss.}}

\documentclass[journal,draftcls,onecolumn,12pt,twoside]{IEEEtran} %TCOM template remember to activate double spaced

\usepackage{multicol}
\usepackage{lipsum}
\usepackage{mathtools}
\usepackage{amsthm}
\usepackage{acronym}
\usepackage{setspace}
\usepackage[bookmarksopen=true]{hyperref}
\usepackage{stfloats}
\usepackage{float}
\usepackage{amsfonts}
\usepackage{acronym}
\usepackage{cite}
\usepackage{multirow}
\usepackage{bm}
\usepackage{amsmath,amssymb}
\usepackage{graphicx}
\usepackage[table,xcdraw]{xcolor}
\usepackage[english]{babel}
\usepackage[caption=false,font=footnotesize]{subfig}
\usepackage[geometry]{ifsym}
\usepackage{array}
\usepackage[utf8]{inputenc}
\usepackage[T1]{fontenc}
\usepackage{graphicx}
\usepackage{lipsum}

\usepackage{tikz}
\usepackage{lipsum}
\usetikzlibrary{calc,shadings,patterns}
\makeatletter
\tikzset{%
  remember picture with id/.style={%
    remember picture,
    overlay,
    save picture id=#1,
  },
  save picture id/.code={%
    \edef\pgf@temp{#1}%
    \immediate\write\pgfutil@auxout{%
      \noexpand\savepointas{\pgf@temp}{\pgfpictureid}}%
  },
  if picture id/.code args={#1#2#3}{%
    \@ifundefined{save@pt@#1}{%
      \pgfkeysalso{#3}%
    }{
      \pgfkeysalso{#2}%
    }
  }
}

\def\savepointas#1#2{%
  \expandafter\gdef\csname save@pt@#1\endcsname{#2}%
}

\def\tmk@labeldef#1,#2\@nil{%
  \def\tmk@label{#1}%
  \def\tmk@def{#2}%
}

\tikzdeclarecoordinatesystem{pic}{%
  \pgfutil@in@,{#1}%
  \ifpgfutil@in@%
    \tmk@labeldef#1\@nil
  \else
    \tmk@labeldef#1,(0pt,0pt)\@nil
  \fi
  \@ifundefined{save@pt@\tmk@label}{%
    \tikz@scan@one@point\pgfutil@firstofone\tmk@def
  }{%
  \pgfsys@getposition{\csname save@pt@\tmk@label\endcsname}\save@orig@pic%
  \pgfsys@getposition{\pgfpictureid}\save@this@pic%
  \pgf@process{\pgfpointorigin\save@this@pic}%
  \pgf@xa=\pgf@x
  \pgf@ya=\pgf@y
  \pgf@process{\pgfpointorigin\save@orig@pic}%
  \advance\pgf@x by -\pgf@xa
  \advance\pgf@y by -\pgf@ya
  }%
}

\makeatother
\newcounter{hatchNumber}
\usepackage{cuted}
\makeatletter
\newif\ifAC@uppercase@first%
\def\Aclp#1{\AC@uppercase@firsttrue\aclp{#1}\AC@uppercase@firstfalse}%
\def\AC@aclp#1{%
	\ifcsname fn@#1@PL\endcsname%
	\ifAC@uppercase@first%
	\expandafter\expandafter\expandafter\MakeUppercase\csname fn@#1@PL\endcsname%
	\else%
	\csname fn@#1@PL\endcsname%
	\fi%
	\else%
	\AC@acl{#1}s%
	\fi%
}%
\def\Acp#1{\AC@uppercase@firsttrue\acp{#1}\AC@uppercase@firstfalse}%
\def\AC@acp#1{%
	\ifcsname fn@#1@PL\endcsname%
	\ifAC@uppercase@first%
	\expandafter\expandafter\expandafter\MakeUppercase\csname fn@#1@PL\endcsname%
	\else%
	\csname fn@#1@PL\endcsname%
	\fi%
	\else%
	\AC@ac{#1}s%
	\fi%
}%
\def\Acfp#1{\AC@uppercase@firsttrue\acfp{#1}\AC@uppercase@firstfalse}%
\def\AC@acfp#1{%
	\ifcsname fn@#1@PL\endcsname%
	\ifAC@uppercase@first%
	\expandafter\expandafter\expandafter\MakeUppercase\csname fn@#1@PL\endcsname%
	\else%
	\csname fn@#1@PL\endcsname%
	\fi%
	\else%
	\AC@acf{#1}s%
	\fi%
}%
\def\Acsp#1{\AC@uppercase@firsttrue\acsp{#1}\AC@uppercase@firstfalse}%
\def\AC@acsp#1{%
	\ifcsname fn@#1@PL\endcsname%
	\ifAC@uppercase@first%
	\expandafter\expandafter\expandafter\MakeUppercase\csname fn@#1@PL\endcsname%
	\else%
	\csname fn@#1@PL\endcsname%
	\fi%
	\else%
	\AC@acs{#1}s%
	\fi%
}%
\edef\AC@uppercase@write{\string\ifAC@uppercase@first\string\expandafter\string\MakeUppercase\string\fi\space}%
\def\AC@acrodef#1[#2]#3{%
	\@bsphack%
	\protected@write\@auxout{}{%
		\string\newacro{#1}[#2]{\AC@uppercase@write #3}%
	}\@esphack%
}%
\def\Acl#1{\AC@uppercase@firsttrue\acl{#1}\AC@uppercase@firstfalse}
\def\Acf#1{\AC@uppercase@firsttrue\acf{#1}\AC@uppercase@firstfalse}
\def\Ac#1{\AC@uppercase@firsttrue\ac{#1}\AC@uppercase@firstfalse}
\def\Acs#1{\AC@uppercase@firsttrue\acs{#1}\AC@uppercase@firstfalse}
\acrodef{CSS}{chirp spread spectrum}
\acrodef{SIC}{successive interference cancellation}
\acrodef{PAPR}{peak-to-average-power ratio}
\acrodef{APAC}{aperiodic autocorrelation}
\acrodef{OFDM}{orthogonal frequency division multiplexing}
\acrodef{DFT}{discrete Fourier transform}
\acrodef{DC}{direct current}
\acrodef{CS}{complementary sequence}
\acrodef{GCP}{Golay complementary pair}
\acrodef{ANF}{algebraic normal form}
\acrodef{PSK}{phase shift keying}
\acrodef{QAM}{quadrature amplitude modulation}
\acrodef{QPSK}{quadrature phase shift keying}
\acrodef{GDJ}{Golay-Davis-Jedwab}
\acrodef{PMEPR}{peak-to-mean envelope power ratios}
\acrodef{FFT}{fast Fourier transform}
\acrodef{BER}{bit-error rate}
\acrodef{SNR}{signal-to-noise ratio}
\acrodef{4G}{Fourth Generation}
\acrodef{5G}{Fifth Generation}
\acrodef{NR}{5G New Radio}
\acrodef{LTE}{Long-Term Evolution}
\acrodef{PTS}{partial transmit sequences}
\acrodef{PSD}{power spectral density}
\acrodef{LDPC}{low-density parity check}
\acrodef{SE}{spectral efficiency}
\acrodef{eLAA}{enhanced licensed-assisted access}
\acrodef{NR-U}{NR-Unlicensed}
\acrodef{RM}{Reed-Muller}
\acrodef{AE}{autoencoder}
\acrodef{DNN}{deep neural network}
\acrodef{OFDM-AE}{OFDM-based autoencoder}
\acrodef{DL}{deep learning}
\acrodef{CP}{cyclic prefix}
\acrodef{AWGN}{additive white Gaussian noise}
\acrodef{P2C}{polar-to-Cartesian}
\acrodef{CFR}{channel frequency response}
\acrodef{ReLU}{rectified linear unit}
\acrodef{MMSE}{minimum mean sqaure error}
\acrodef{BPSK}{binary phase shift keying}
\acrodef{BLER}{block error rate}
\acrodef{ML}{machine learning}
\acrodef{PHY}{physical layer}
\acrodef{PA}{power amplifier}
\acrodef{IDFT}{inverse DFT}
\acrodef{DoF}{degrees-of-freedom}
\acrodef{IoT}{Internet-of-Things}
\acrodef{DFT-s-OFDM}{discrete Fourier transform-spread orthogonal frequency division multiplexing}
\acrodef{MMSE}{minimum mean square error}
\acrodef{FDE}{frequency-domain equalization}
\acrodef{FrFT}{fractional Fourier transform}
\acrodef{TF}{time-frequency}
\acrodef{BFSK}{binary frequency shift keying}
\acrodef{CSS}{chirp spread spectrum}
\acrodef{BCSS}{binary chirp spread spectrum}
\acrodef{EVA}{Extended Vehicular A}
\acrodef{MIMO}{multi-input multi-output}
\acrodef{PIC}{parallel interference cancellation}
\acrodef{LoRa}{Long Range}
\acrodef{HF}{high-frequency}
\begin{document}
\title{ 
Noncoherent Multiuser Chirp Spread Spectrum: Performance with Doppler and Asynchronism
}
%Non-contiguous Complementary Sequence Encoder
%A Generalized Complementary Sequence Encoder
\author{Nozhan~Hosseini,~\IEEEmembership{Member,~IEEE},  David~W.~Matolak,~\IEEEmembership{Senior Member,~IEEE} 
\thanks{This work was partially supported by NASA, under award number
NNX17AJ94A. The authors are with the University of South Carolina, Columbia, SC. E-mail: nozhan@cec.sc.edu, matolak@cec.sc.edu.
}

}
\maketitle
\doublespacing
\markboth{IEEE Transactions on Communications}%
{Submitted paper}
\begin{abstract}
%In this study, we propose a linear and sinusoidal chirp-based communication system by utilizing  Bessel functions and Fresnel integrals through \ac{DFT-s-OFDM}. This scheme offers a way to efficiently generate chirp signals that can be used in Internet of Things or radar applications with existing \ac{DFT-s-OFDM} transceivers.
In this paper, we investigate multi user chirp spread spectrum with noncoherent detection as a continuation of our work on coherent detection in \cite{journal1}. We derive the analytical bit error ratio (BER) expression for binary chirp spread spectrum (BCSS) in the presence of multiple access interference (MAI) caused by correlation with other user signals because of either asynchronism or Doppler shifts, or both, and validate with simulations. To achieve this we analyze the signal cross correlations, and compare traditional linear chirps with our recently-proposed nonlinear chirps introduced in \cite{journal1} and with other nonlinear chirps from the literature. In doing so we illustrate the superior performance of our new nonlinear chirp designs in these practical conditions, for the noncoherent counterpart of \cite{journal1}.
\end{abstract}
\acresetall
\section{Introduction}\label{Introduction}
Communication systems experience multiple impairments depending on their environment. These include multipath channel distortion, Doppler spreading, and interference. Nonlinear distortion due to equipment (e.g., high power amplifier) non-idealities are also present, and these are particularly challenging for commonly used multicarrier signals, and even for single-carrier signals that employ non-constant-envelope signaling, e.g., amplitude modulation such as in \ac{QAM}. Thus other signal types are of interest, and frequency modulated signals such as chirps are one such signal type. Chirps have been investigated for multiple purposes, e.g., communications, radar and channel characterization \cite{Nozhan_wideband_sounding,Radar,radar2,Communication_litreture,communication_lit2,nozhan_conf}. 

Constant amplitude chirp signals exhibit a desirable low \ac{PAPR} which enables longer link range or use of less expensive amplifiers. Chirps also have a sharp ridge-like autocorrelation, and this enables chirp signals to be used for channel characterization measurements (sounding) which can lead to a hybrid communication/sounding system. A disadvantage of chirps is that they generally require excess bandwidth, and are hence spread spectrum (SS) signals. To alleviate the low spectral efficiency of chirp SS (CSS) requires multiple user signals be supported within a band. One challenge with this is inter-signal interference, or multiple-access interference (MAI), which degrades performance. There are well-known CSS schemes that can eliminate MAI, but these generally require perfect synchronization and minimal Doppler shifts. Channel dispersion can also induce MAI (as well as inter-symbol interference, ISI). 

Most CSS receivers are coherent \cite{journal1,ouyang1,coherent_chirp_alsharef,coherent_wang,CSS_QIAN,CSS_NGUYEN}. Although coherent detection offers better peformance (bit error ratio, BER) than non-coherent, in some channels phase estimation is difficult. In addition, phase estimation adds to circuit complexity and cost. Expensive coherent receivers usually use phase-locked loop circuits to recover the carrier phase of the received signal, whereas cheaper noncoherent receivers require zero knowledge of received signal phase. Noncoherent receivers treat the unknown phase as a random variable and in a sense average the likelihood of the received signal with unknown phase with respect to the distribution of the phase (typically well-modeled as uniform). Thus in addition to theoretical interest in noncoherent detection, such schemes are also of interest in lower-cost applications and when BER performance is less critical.

In this paper, first, for reference, in Section \ref{Coherent Detection} we provide a very brief description of coherent detection BER from \cite{journal1}. We then analyze non-coherent detection for multi-user CSS, which to the best of our knowledge is investigated for the first time in Section \ref{Noncoherent Detection}. Imposing asynchronism and Doppler shifts in Section \ref{sec:Doppler and Asynchronous Effect}, we show inter-signal cross correlation results and BER performance results from both analysis and simulations. Section \ref{sec:conclusion} concludes the paper.

\section{Coherent Detection}\label{Coherent Detection}
In \cite{journal1}, we described coherent detection in detail. Here we only provide the derived BER expression from \cite{journal1}, which we later use in presenting results. This equation pertains to binary CSS in which binary symbol zero is represented by an "upchirp" in a frequency band of width $2N/T$, and symbol one is represented by an upchirp in a directly adjacent frequency band of the same width. (Results also pertain to "downchirps.") In this arrangement, the individual user chirp signals represent a form of frequency-shift keying (FSK). Here $N$ is the total number of user signals, and $T$ is the symbol (bit) duration. The BER, for the kth user signal in $any$ type of BCSS signal set, is,
\begin{equation}
\begin{aligned}
    P_{b,k} = \frac{1}{2^{N-1}}\sum_{\zeta=0}^{2^{N-1}-1}Q\left(\sqrt{\frac{(1+\pmb{\rho}_k^T\pmb{b}_\zeta)^2E_{sk}}{N_0}} \right)
    \label{COHERENT_EQUATION_BER}
    \end{aligned}
\end{equation}
where the Q-function is the tail integral of the zero-mean, unit variance Gaussian density function, \textcolor{black}{ $E_{sk}$ is the symbol (bit) energy and} $\pmb{b}_\zeta$ is a vector of size $(N-1)\times1$, with elements in the set $\{-1,1\}$ defined as,
\begin{equation}
    \pmb{b}_{\xi}=\left[\begin{matrix}(-1)^{a\left(\xi,0\right)}\ \\(-1)^{a\left(\xi,1\right)}\\\vdots\\(-1)^{a\left(\xi,N-2\right)}\\\end{matrix}\right]
    \label{eq:noncoherent(dummy)}.
\end{equation}
 Variable $a\left(\xi,i\right)\in\left\{0,1\right\}$ is the $i$th coefficient in the binary expansion for decimal number $\xi$, i.e,
\begin{equation}
    \xi=\sum_{i=0}^{N-1} a(\xi,i)2^i
    \label{eq:noncoherent34}
\end{equation}
and $\pmb{\rho}_{k}$ is the cross correlation vector of dimension $(N-1)\times1$, which is a column of the complete correlation matrix. This vector is

\begin{equation}
    \pmb{\rho}_{k}=\pmb{\rho}_{km}\setminus\pmb{\rho}_{kk}=\left[\begin{matrix}\sqrt{\frac{E_{s0}}{{E}_{{sm}}}}\rho_{0,m}\ \\\sqrt{\frac{E_{s1}}{{E}_{{sm}}}}\rho_{1,m}\\\vdots\\\sqrt{\frac{E_{sN-1}}{{E}_{{sm}}}}\rho_{N-1,m}\\\end{matrix}\right],
    \label{eq:noncoherent36}
\end{equation}
i.e., vector $\pmb{\rho}_{k}$ includes all cross-correlation values $\rho_{km}$ except $\rho_{kk}$. As one may observe, this BER also allows for arbitrary received signal energies at the single-user chirp receiver.

Considering the chirp signal expression from \cite{journal1} and adding the unknown phase component $\vartheta$ we can write, 
\begin{equation}
\begin{aligned}
    s_m(t) = e^{\frac{j \pi N}{T^2}(t+\frac{mT}{N})^2 + j \vartheta}~~,~~  0\leq t<T
    \end{aligned}
\end{equation}
where $\vartheta$ is the unknown phase, $N$ is the number of users in a set, $T$ is bit duration and $m$ is the user index.

\section{Noncoherent Detection}\label{Noncoherent Detection}
\begin{figure}[ht]
	\centering
	{\includegraphics[width =\linewidth]{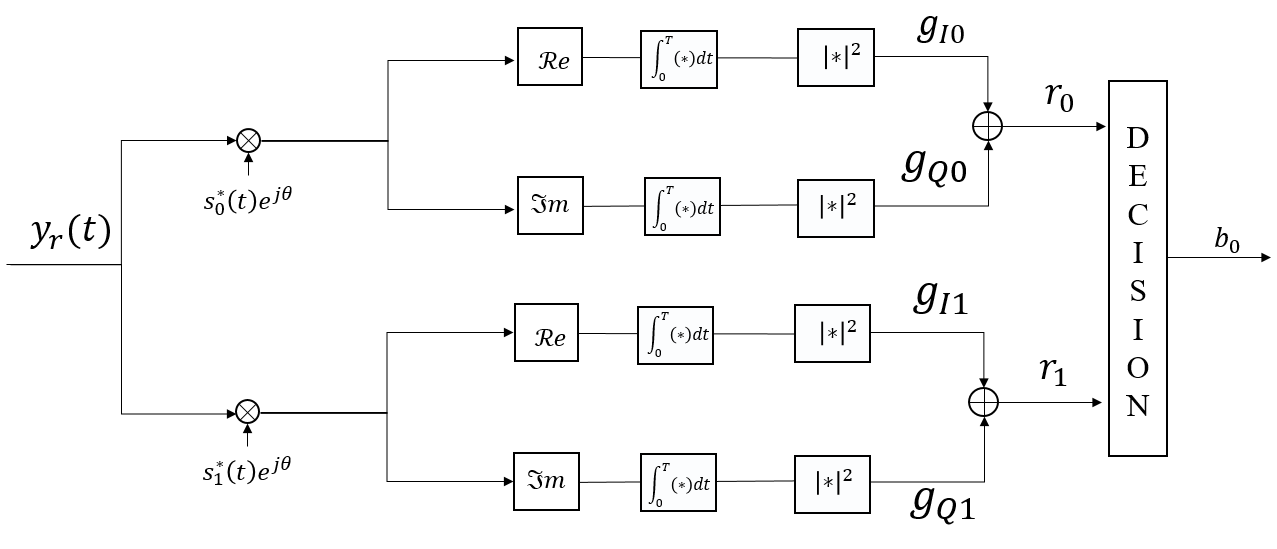}
	}
	\vspace{-3mm}
	\caption{Binary chirp non-coherent detection receiver baseband block diagram.}
	\label{fig:noncoherent_blockdiagram}
\end{figure}

In this section, we investigate non-coherent detection for \ac{CSS} signals. As noted, noncoherent detection is of interest when receiver hardware cannot ensure the local waveforms have the same starting phase as the received signals, e.g., for very inexpensive receivers. The analysis pertains explicitly to the linear chirp case for ease of illustration, but is equally valid for other chirp signals, as we will show. We also assume the additive white Gaussian noise channel, but will incorporate asynchronism, Doppler, and unequal received signal energies. Other channels (e.g., dispersive) are left for future work, but our model is applicable to several practical settings, such as many air-ground channels. The transmitter structure is identical to the coherent case but the noncoherent receiver has a different structure, shown in Fig. \ref{fig:noncoherent_blockdiagram}.

The receiver essentially correlates with the two possible transmitted signals, $s_0$ and $s_1$. The phase $\theta$ represents the unknown receiver phase.

\subsection{Single User BER Analysis}
For our binary linear chirp modulation, the two possible transmitted signals for user $i$ are defined as,
\begin{equation}
\begin{array}{c}
s_{i}(t)=\left\{\begin{array}{ll}
s_{0}(t)=A  e^{\frac{j \pi N}{T^{2}} t^{2}} & \text { if "0" transmitted } \\
s_{1}(t)=A  e^{\frac{j \pi N}{T^{2}}(t+T)^{2}} & \text { if "1" transmitted }
\end{array}\right.
\quad ,0<t<T.
\end{array}
\label{eq:noncoherent0}
\end{equation}
Here, $A$ is the transmitted signal amplitude and $N/T^2$ is the chirp rate. Considering the unknown phase at the receiver for non-coherent detection, to assess performance we must find the variables $g_{I0}$,\ \ $g_{Q0}$ and $g_{I2}$,\ \ $g_{Q2}$,\, shown in Fig. \ref{fig:noncoherent_blockdiagram}, when our user sends “0”. (Via symmetry, performance when a “1” is sent is identical.) Since the channel is AWGN, performance can be evaluated via single-symbol detection. According to Fig. \ref{fig:noncoherent_blockdiagram} we can write,

\begin{equation}
\begin{aligned}
g_{I 0}=&\left|\left.\int_{0}^{T} \Re e\left\{\left[A  e^{\frac{j \pi N}{T^{2}} t^{2}}+w(t)\right]  e^{-\frac{j \pi N}{T^{2}} t^{2} +j \theta}\right\}\right|^{2}\right.,\\
g_{Q 0}=&\left|\left.\int_{0}^{T} \Im m\left\{\left[A  e^{\frac{j \pi N}{T^{2}} t^{2}}+w(t)\right]  e^{-\frac{j \pi N}{T^{2}} t^{2}+j \theta}\right\}\right|^{2}\right.,\\ \\
g_{I 1}=&\left| \int_{0}^{T} \Re e\left\{\left[A  e^{\frac{j \pi N}{T^{2}} t^{2}}+w(t)\right] e^{-\frac{j \pi N}{T^{2}}(t+T)^{2}+j \theta}\right\}\right|^{2} ,\\
g_{Q 1}=&\left|\int_{0}^{T} \Im m\left\{\left[A  e^{\frac{j \pi N}{T^{2}} t^{2}}+w(t)\right]  e^{-\frac{j \pi N}{T^{2}}(t+T)^{2}+j \theta}\right\}\right|^{2},
\label{eq:noncoherent1}
\end{aligned}
\end{equation}
where $w\left(t\right)$ is the white Gaussian noise. With some simplifications we can obtain,
\begin{equation}
\begin{array}{c}
g_{I 0}=\left(A T \cos (\theta)+n I_{0}\right)^{2} \\
g_{Q 0}=\left(A T \sin (\theta)+n Q_{0}\right)^{2} \\ \\ 
g_{I 1}=\left(n I_{1}\right)^{2} \\
g_{O 1}=\left(n Q_{1}\right)^{2}
\label{eq:noncoherent2}
\end{array}
\end{equation}
where $nI_0$, $nQ_0$, $nI_1$ and $nQ_1$  are jointly Gaussian independent random variables with zero means and defined as,
\begin{equation}
    \begin{array}{l}
n I_{v}=\int_{0}^{T} w_{\Re}(t) s_{v}(t) e^{j \theta} \\
n Q_{v}=\int_{0}^{T} w_{\Im}(t) s_{v}(t) e^{j \theta}
\label{eq:noncoherent3}
\end{array}
\end{equation}
with $v$ equal to 0 or 1. The effect of phase uncertainty $e^{j\theta}$ can be absorbed into the noise component, whose distribution is circularly symmetric, hence phase rotation will not affect the statistics. The variance of the components in \eqref{eq:noncoherent3}, using our assumption $N/T >>1$ can be found as,
\begin{equation}
    E\left[n I^{2}\right]=E\left[n Q^{2}\right] \cong N_{0} T.
\label{eq:noncoherent4}
\end{equation}
The decision statistic $r_1$ is then given by,
\begin{equation}
r_{1}=\sqrt{\left(n I_{1}\right)^{2}+\left(n Q_{1}\right)^{2}}.
\label{eq:noncoherent5}
\end{equation}
We know that the square root of the sum of the squares of two independent Gaussian random variables with both zero mean and the same variance is a Rayleigh random variable, with density function given by,
\begin{equation}
f_{r_{1}}=\frac{r_{1}}{\sigma^{2}} \exp \left(-\frac{r_{1}^{2}}{2 \sigma^{2}}\right)
\label{eq:noncoherent6}
\end{equation}
where $\sigma^2\cong N_0 T$.
The Rician distribution is the result of square root of the sum of squares of two independent and identically distributed Gaussian random variable with the same $\sigma$,  and non-zero means,
\begin{equation}
f_{x}=\frac{x}{\sigma^{2}} e^{-\frac{x^{2}+a^{2}}{2 \sigma^{2}} I_{0}\left(\frac{a x}{\sigma^{2}}\right)}
\label{eq:noncoherent7}
\end{equation}
where $a=\sqrt{\mu_{1}^{2}+\mu_{2}^{2}}=\sqrt{A^{2} T^{2} \cos ^{2}(\theta)+A^{2} T^{2} \sin ^{2}(\theta)}=A T$. This pertains to the decision statistic $r_0$, whose pdf can be written as,
\begin{equation}
f_{r_{0}}=\frac{r_{0}}{\sigma^{2}} \exp \left(-\frac{r_{0}^{2}+A^{2} T^{2}}{2 \sigma^{2}}\right) I_{0}\left(\frac{A T r_{0}}{\sigma^{2}}\right)
\label{eq:noncoherent8}
\end{equation}
where $I_0$ is the modified Bessel function of first kind, zero order.\\

These two distributions are independent $f_{R_{1}, R_{0}}\left(r_{1}, r_{0}\right)=f\left(r_{0}\right) f\left(r_{1}\right)$, and we can write the joint pdf as the product of \eqref{eq:noncoherent8} and \eqref{eq:noncoherent6}. Then with our assumption of a zero sent, the probability of error is,
\begin{equation}
\begin{aligned}
     P\left(r_{1}>r_{0}\right)& =\int_{0}^{\infty} d r_{0} \int_{r_{0}}^{\infty} f\left(r_{1}, r_{0}\right) d r_{1} %\\ 
     =\int_{0}^{\infty} f_{R_{0}\left(r_{0}\right)}\left(\int_{r_{0}}^{\infty} f_{R 1}\left(r_{1}\right) d r_{1}\right) d r_{0} \\ 
 &\left.=\int_{0}^{\infty} f_{R_{0}\left(r_{0}\right)}\left(-\exp \left(-\frac{r_{1}^{2}}{2 \sigma^{2}}\right)\right]_{r_{0}}^{\infty}\right) d r_{0}%\\ 
 =\int_{0}^{\infty} \exp \left(\frac{r_{0}^{2}}{2 \sigma^{2}}\right) f_{R_{0}}\left(r_{0}\right) d r_{0}\\
& =\int_{0}^{\infty} \frac{r_{0}}{\sigma^{2}} \exp \left(-\frac{2 r_{0}^{2}+A^{2} T^{2}}{2 \sigma^{2}}\right) I_{0}\left(\frac{A T r_{0}}{\sigma^{2}}\right) d r_{0}.
 \end{aligned}
 \label{eq:noncoherent10}
\end{equation}

Now if we change variables, $u=\sqrt{2}r_0$ and $q=\frac{AT}{\sqrt{2}}$, we have
\begin{equation}
\begin{aligned}
P\left(r_{1}>r_{0}\right)& =    \int_{0}^{\infty} \frac{u}{\sqrt{2} \sigma^{2}} \exp \left(-\frac{u^{2}+2 q^{2}}{2 \sigma^{2}}\right) I_{0}\left(\frac{q u}{\sigma^{2}}\right) \frac{d u}{\sqrt{2}} \\
 & =\frac{1}{2} e^{-\frac{q^{2}}{2 \sigma^{2}}} \int_{0}^{\infty} \frac{u}{\sigma^{2}} \exp \left(-\frac{u^{2}+q^{2}}{2 \sigma^{2}}\right) I_{0}\left(\frac{q u}{\sigma^{2}}\right) d u.
\end{aligned}
 \label{eq:noncoherent11}
\end{equation}
The integral in \eqref{eq:noncoherent11} is the integral of the Rician density, and hence is equal to one. and therefore,
\begin{equation}
P_{b}(0~sent)=\frac{1}{2} \exp \left(-\frac{q^{2}}{2 \sigma}\right)=\frac{1}{2} \exp \left(-\frac{A^{2} T^{2}}{4 \sigma^{2}}\right).
 \label{eq:noncoherent12}
\end{equation}
The average error probability is obtained by averaging this expression over the random unknown phase distribution. We model this phase as uniformly distributed, hence the average error probability when a zero is sent can be written as,

\begin{equation}
\begin{aligned} P_{b_{A v e}}(0~s e n t) &=\frac{1}{2 \pi} \int_{0}^{2 \pi} \frac{1}{2} \exp \left(-\frac{A^{2} T^{2}}{4 \sigma^{2}}\right) d \theta \\ &=\frac{1}{2} \exp \left(-\frac{A^{2} T^{2}}{4 \sigma^{2}}\right)=\frac{1}{2} \exp \left(-\frac{2 E_{s} T}{4 \sigma^{2}}\right)\\
&=\frac{1}{2} \exp \left(-\frac{E_{s}}{2 N_{0}}\right) \end{aligned}
 \label{eq:noncoherent13}
\end{equation}
which agrees with the result for NC detection of binary FSK, as it should. Note again that $P_b (0~sent)=P_b (1~sent)$=$P_{b,avg}$.
\subsection{Multi User Asynchronous BER Performance}

\begin{figure*}[h!]
	\centering
	{\includegraphics[width =\linewidth]{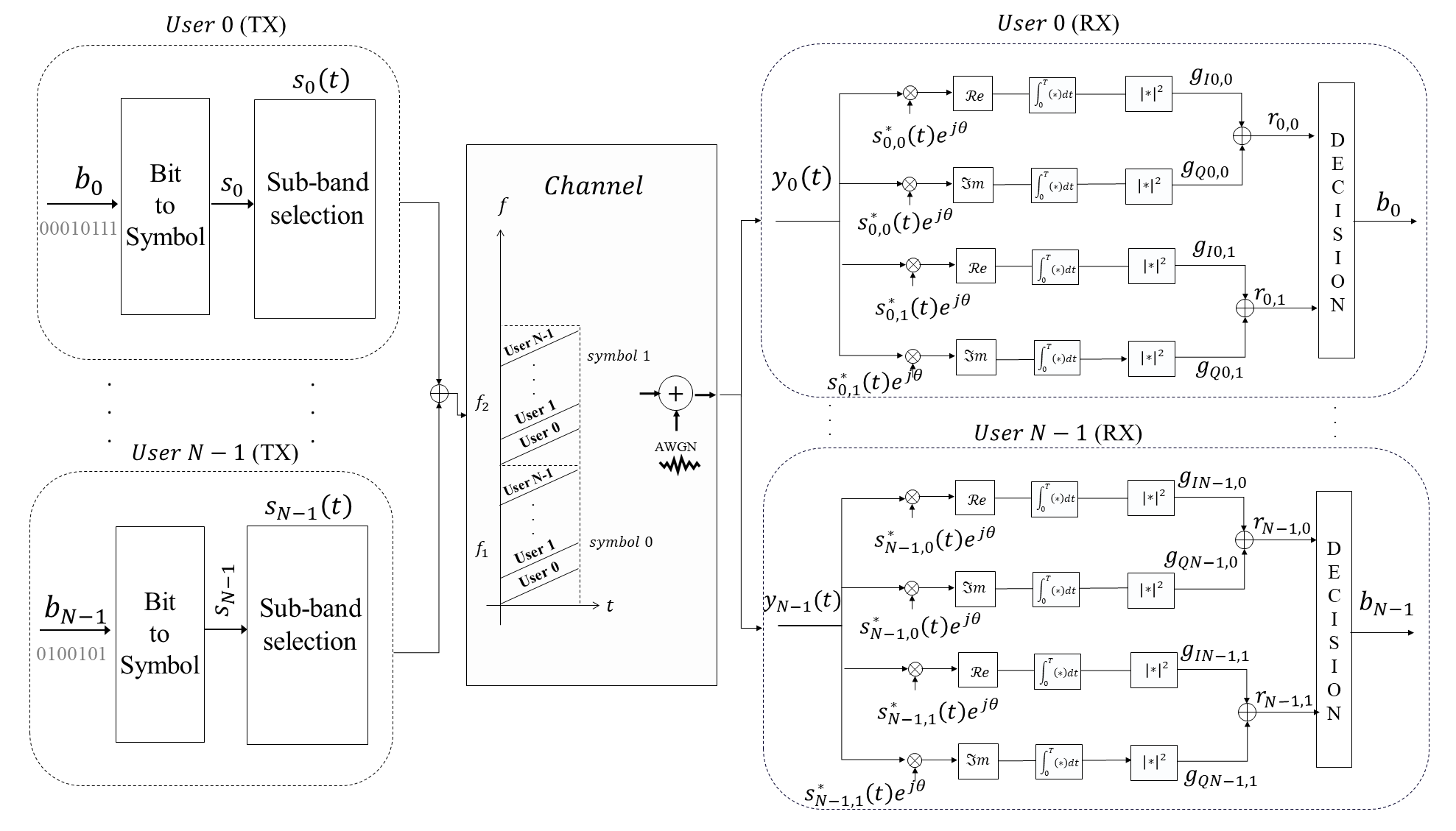}
	} 
	\caption{General multiuser, $M$-ary non-coherent CSS detection block diagram.}
	\vspace{-3mm}
	\label{fig:Multi_user_block_diagram}
\end{figure*}

As noted in the Introduction, many CSS signal sets can employ orthogonal signals if synchronization is perfect at receivers. In such a case, there is no MAI, and BCSS performance for all signals is that for the single user case analyzed in the prior section. In the practical case where perfect synchronization is not possible, the analysis must quantify and incorporate MAI. Fig. \ref{fig:Multi_user_block_diagram} shows a block diagram of a multi-user noncoherent BCSS system for $N$ users. The signals output from the correlators of Fig. \ref{fig:Multi_user_block_diagram} are similar to those described in the previous section, but now we assume multiple users.
 We first analyze the two-user case and then generalize to an arbitrary number of users. In the two-user asynchronous case, user one’s signal is delayed by $\epsilon$. We assume that user 0 again sends a symbol "0". In this case the correlator “g” variables of Fig. \ref{fig:Multi_user_block_diagram} are,
\begin{equation}
\begin{aligned}
\begin{rcases}
g_{I v}&=\left|\int \Re e\left\{\left[s_{00}(t)+s_{10}(t-\epsilon)+w(t)\right] s_{00}^{*}(t) e^{j \theta}\right\}\right|^{2} \\
g_{Q v}&=\left|\int \Im m\left\{\left[s_{00}(t)+s_{10}(t-\epsilon)+w(t)\right] s_{00}^{*}(t) e^{j \theta}\right\}\right|^{2}  \end{rcases} user1~sent~"0" \\ \\
\begin{rcases}g_{I v}&=\left|\int \Re e\left\{\left[s_{00}(t)+s_{11}(t-\epsilon)+w(t)\right] s_{01}^{*}(t) e^{j \theta}\right\}\right|^{2} \\
g_{Q v}&=\left|\int \Im m\left\{\left[s_{00}(t)+s_{11}(t-\epsilon)+w(t)\right] s_{01}^{*}(t) e^{j \theta}\right\}\right|^{2} \end{rcases} user1~sent~"1"
\label{eq:noncoherent14}
\end{aligned}
\end{equation}
In the fully synchronized case, it is well known that any two such linear \ac{CSS} signals are orthogonal. In case of asynchronism (or, quasi-synchronism) represented by delay $\epsilon$, orthogonality is violated and the user signals have a non-zero cross correlation, inducing MAI. The cross correlation is,
\begin{equation}
\begin{aligned}
\rho_{01}(\epsilon)&=\int\left[s_{10}(t-\epsilon) s_{00}^{*}(t) e^{j \theta}\right]%\\
=\int\left[A e^{\frac{j \pi N}{T^{2}}\left(t+\frac{T}{N}-\epsilon\right)^2} \times A e^{-\frac{j \pi }{4}} e^{-\frac{j \pi N}{T^{2}} t^{2} j \theta}\right]\\
&=A T[\rho_{10_\Re} \cos (\theta)-\rho_{10_\Im} \sin(\theta) +i(\rho_{10_\Im} \cos (\theta)%\\
+\rho_{10_\Re}  \sin (\theta))].
\label{eq:noncoherent15}
\end{aligned}
\end{equation}
The two signals are correlated with complex cross correlation ($\rho_\Re+j\rho_\Im$ ).  
where $\rho_{10_\Re}$ and $\rho_{10_\Im}$ are the real and imaginary parts of the cross correlation between user 1’s signal delayed by $\epsilon$ relative to that of user 0’s signal. The expanded decision variables, when user 0 sends “0,” written in \eqref{eq:noncoherent16}.

\begin{equation}
\begin{aligned}
\begin{rcases}
g_{I 0,0}&=\left| AT\operatorname{cos}(\theta)+A T \rho_{10_\Re} \cos (\theta)-A T \rho_{10_\Im} \sin (\theta)
+\int \Re e\left\{w(t) s_{00}^{*}(t) e^{j \theta}\right.\}\right|^2 \\
g_{Q 0,0}&=\left| AT \operatorname{sin}(\theta)+A T \rho_{10_\Im} \cos (\theta)+A T \rho_{10_\Re} \sin (\theta)
+\int \Im m\left\{w(t) s_{00}^{*}(t) e^{j \theta}\right\}\right|^2 \\
g_{I0,1}&=\left|\int \Re e\left\{w(t) s_{01}^{*}(t) e^{j \theta}\right\}\right|^{2} \\
g_{Q 0,1}&=\left|\int \Im m\left\{w(t) s_{01}^{*}(t) e^{j \theta}\right\}\right|^{2}  \end{rcases} i \\ \\
\begin{rcases}
g_{I 0,0}&=\left| AT\operatorname{cos}(\theta)
+\int \Re e\left\{w(t) s_{00}^{*}(t) e^{j \theta}\right.\}\right|^2 \\
g_{Q 0,0}&=\left| AT \operatorname{sin}(\theta)
+\int \Im m\left\{w(t) s_{00}^{*}(t) e^{j \theta}\right\}\right|^2. \\
g_{I0,1}&=\left|A T \rho_{10_\Re} \cos (\theta)-A T \rho_{10_\Im} \sin (\theta)+\int \Re e\left\{w(t) s_{01}^{*}(t) e^{j \theta}\right\}\right|^{2} \\
g_{Q 0,1}&=\left|+A T \rho_{10_\Im} \cos (\theta)+A T \rho_{10_\Re} \sin (\theta)+\int \Im m\left\{w(t) s_{01}^{*}(t) e^{j \theta}\right\}\right|^{2}  \end{rcases} ii
\label{eq:noncoherent16}
\end{aligned}
\end{equation}
where $i$ and $ii$ denote if user one sent "0" and "1" respectively.
Statistically, we assume each symbol is equiprobable. We continue without loss of generality to assume that user zero sends “0”. Following the same procedure as in the single user fully synchronized case, for user one sending a “0,” since the two branch variables  $g_{I_{0,1}}$ and $g_{Q_{0,1} }$ are added, as described in \eqref{eq:noncoherent5}-\eqref{eq:noncoherent6}, this leads to a  Rayleigh distribution for the noise-only variables. Here $a$ in \eqref{eq:noncoherent7} can be found from \eqref{eq:noncoherent15} as,
\begin{equation}
\begin{aligned}
    a&=\sqrt{\mu_{1}^{2}+\mu_{2}^{2}}%\\
    =\sqrt{\begin{multlined}[b]  \begin{array}{l}
(AT\operatorname{cos}(\theta)+AT \rho_{10_\Re} \cos (\theta)-AT \rho_{10_\Im} \sin (\theta))^{2} \\
+\left(AT\operatorname{sin}(\theta)+AT \rho_{10_\Im} \cos (\theta)+A T \rho_{10_\Re} \sin (\theta)\right)^{2}\end{array} \end{multlined}} \\
&=AT\sqrt{1+\rho_{10_\Re}^2+\rho_{10_\Im}^2+2\rho_{10_\Re}}
\label{eq:noncoherent17}
\end{aligned}
\end{equation}
where $\mu_1$ and $\mu_2$ are the mean values of $g_{I0,0}$ and $g_{Q0,0}$, respectively. By substituting \eqref{eq:noncoherent15} into \eqref{eq:noncoherent7} and following the development of \eqref{eq:noncoherent8} to \eqref{eq:noncoherent13}, we can find the final error probability expression when user 0 sends “0” as,

\begin{equation}
\begin{aligned}
   & P_{b}(user 1~sent"0") =0.5 \exp \left(-\frac{E_{s} \sqrt{1+\rho_{10_\Re}^2+\rho_{10_\Im}^2+2 \rho_{10_\Re}}}{2 N_{0}}\right).
\label{eq:noncoherent18}
\end{aligned}
\end{equation}
We note that if the correlations are zero, i.e., for the synchronous case, \eqref{eq:noncoherent18} reduces to the single-user (NC FSK) result, as expected.\par
Next, if user 1 sends “1”, the Rician distribution again results for the RV formed as described in the single user case by the square root of the sum of squares of two independent ($g_{I_{00} }$ and $g_{Q_{00} }$) and identically distributed Gaussian random variables with the same $\sigma$ but different mean value $\mu$. This Rician density is

\begin{equation}
\begin{aligned}
   p(r_{0}|X,\sigma) =\frac{r}{\sigma^2} \exp \left(-\frac{r_{0}^2+X^2}{2\sigma^2}\right)I_{0}\left(\frac{X r_{0}}{\sigma^2}\right)
\label{eq:noncoherent19}
\end{aligned}
\end{equation}
where $X=\sqrt{\left(ATcos(\theta)\right)^2+\left(ATsin(\theta)\right)^2}=AT$
\par
We use the same procedure as for the single-user case, but with the following densities,
\begin{equation}
\begin{aligned}
   f_{r_{0}} =\frac{r_{0}}{\sigma^2} \exp \left(-\frac{r_{0}^2+X^2}{2\sigma^2}\right)I_{0}\left(\frac{X r_{0}}{\sigma^2}\right)
\label{eq:noncoherent20}
\end{aligned}
\end{equation}
\begin{equation}
\begin{aligned}
   f_{r_{1}} =\frac{r_{1}}{\sigma^2} \exp \left(-\frac{r_{1}^2+Y^2}{2\sigma^2}\right)I_{0}\left(\frac{Y r_{1}}{\sigma^2}\right)
\label{eq:noncoherent21}
\end{aligned}
\end{equation}
where
\begin{equation}
\begin{aligned}
Y &=\left(\begin{multlined}[b]
\left(AT\rho_{10_{\Re}}cos(\theta)-
AT\rho_{10_{\Im}}sin(\theta)\right)^2)\\
+\left(AT\rho_{10_{\Im}}cos(\theta)+
AT\rho_{10_{\Re}}sin(\theta)\right)^2\end{multlined}\right)^{\frac{1}{2}}%\\
=AT\sqrt{\rho_{10_{\Re}}^2+\rho_{10_{\Im}}^2}
\end{aligned}
\label{eq:noncoherent22}
\end{equation}
\par
The two distributions are independent so we can write the joint density function as,
\begin{equation}
\begin{aligned}
&f_{R_{1},R_{0}}(r_1,r_0)=\frac{r_{0}r_{1}}{\sigma^4}\exp{\left(-\frac{r_{1}^2+Y^2}{2\sigma^2}\right)}\\
&\times\exp{\left(-\frac{r_{0}^2+X^2}{2\sigma^2}\right)}I_{0}\left(\frac{Xr_{0}}{\sigma^2}\right)I_{0}\left(\frac{Yr_{1}}{\sigma^2}\right) for~r_{1},r_{0} \geq ~0.
\end{aligned}
\label{eq:noncoherent23}
\end{equation}
Then the error probability expression for this case is,
\begin{equation}
\begin{aligned}
&P(r_{1} >r_{0})=\int_0^\infty dr_{0}\int_0^\infty f(r_{1},r_{0})dr_{1}%\\
=\int_0^\infty f_{R_{0}}(r_0)\left(\int_{r_{0}}^\infty f_{R_{1}}(r_{1})dr_{1}\right)dr_{0}\\
&=\int_0^\infty f_{R_{0}}(r_0)\int_{r_{0}}^\infty \frac{r_{1}}{\sigma^2}\exp{\left(-\frac{r_{1}^2+Y^2}{2\sigma^2}\right)}I_{0}\left(\frac{Yr_{1}}{\sigma^2}\right)dr_{1}r_{0}.
\end{aligned}
\label{eq:noncoherent24}
\end{equation}

\par
Here we find that this integral of the Rician distribution results in a Marcum Q-function, i.e., 
\begin{equation}
\begin{aligned}
\int_n^\infty \exp{\left(-\frac{x^2+m^2}{2\sigma^2} \right)}I_{0}\left(\frac{mx}{\sigma^2} \right)=Q_{1}\left(\frac{m}{\sigma},\frac{n}{\sigma} \right).
\end{aligned}
\label{eq:noncoherent25}
\end{equation}
The Marcum Q-function $Q_k$ is defined as, 

\begin{equation}
\begin{aligned}
Q_{k}(a,b)=\int_b^\infty x\left(\frac{x}{a} \right)^{k-1}\exp{\left[-\frac{x^2+m^2}{2} \right]}I_{k-1}(ax)dx
\end{aligned}
\label{eq:noncoherent26}
\end{equation}

Therefore, we obtain
\begin{equation}
\begin{aligned}
&P(r_{1} >r_{0})=\int_0^\infty f_{R_{0}}(r_{0})Q_{1}\left(\frac{Y}{\sigma},\frac{r_{0}}{\sigma} \right)dr_{0}\\
&=\int_0^\infty \frac{r_{0}}{\sigma^2}\exp{\left(-\frac{r_0^2+X^2}{2\sigma^2} \right)I_0\left(\frac{Xr_{0}}{\sigma^2} \right)}Q_{1}\left(\frac{Y}{\sigma},\frac{r_{0}}{\sigma} \right)dr_{0}\\
&=\frac{1}{\sigma^2}\exp{\left(-\frac{X^2}{2\sigma^2} \right)} \int_0^\infty r_{0} \exp{\left(-\frac{r_{0}^2}{2\sigma^2} \right)}I_{0}\left(\frac{Xr_{0}}{\sigma^2} \right)%\\ &
\times Q_{1}\left(\frac{Y}{\sigma},\frac{r_{0}}{\sigma} \right)dr_{0}
\end{aligned}
\label{eq:noncoherent27}
\end{equation}

The solution for this integral is provided in Appendix as,
\begin{equation}
\begin{aligned}
I&=\int_0^\infty x\exp{\left(-\frac{p^2x^2}{2}\right)}I_{0}(cx)Q(\beta,ax)dx\\
&=\frac{1}{p^2}\left[\begin{multlined}\exp{\left(\frac{c^2}{2p^2}\right)Q\left(\frac{\beta p}{\sqrt{p^2+a^2}},\frac{ac}{p\sqrt{p^2+a^2}} \right) }\\ -\frac{a^2}{p^2+a^2}\exp{\left(\frac{-\beta^2p^2+c^2}{2(p^2+a^2)}\right)}I_{0}\left(\frac{abc}{p^2+a^2} \right)\end{multlined} \right]
\end{aligned}
\label{eq:noncoherent28}
\end{equation}
where here we have $p=\frac{1}{\sigma}$,$c=\frac{X}{\sigma^2}$, $\beta=\frac{Y}{\sigma}$ and $a$=$\frac{1}{\sigma}=p$ for our case and $\sigma=\sqrt{N_0 T}$.

The BER expression when user 1 sends “1” can then be written as,
\begin{equation}
\begin{aligned}
P_{b|1}=Q\left(\frac{Y}{\sigma\sqrt{2}},\frac{X}{\sigma\sqrt{2}} \right)-\frac{1}{2}\exp{\left(-\frac{X^2+Y^2}{4\sigma^4}\right)}I_{0}\left(\frac{XY}{2\sigma^2} \right)
\end{aligned}
\label{eq:noncoherent29}
\end{equation}
\par
Then, the final BER result for two asynchronous (correlated) user signals, noncohrerently detected, can be written by combining \eqref{eq:noncoherent29} and \eqref{eq:noncoherent18} with equal probability as,

\begin{equation}
\begin{aligned}
&P_{b}=\frac{1}{4}\exp{\left(-\frac{E_{s}\sqrt{1+\rho_{10_{\Re}}^2+\rho_{10_{\Im}}^2+2\rho_{10_{\Re}}}}{2N_{0}} \right)}\\
&+\frac{1}{2} \left[Q\left(\frac{Y}{\sigma\sqrt{2}},\frac{X}{\sigma\sqrt{2}} \right)-\frac{1}{2} \exp{\left(-\frac{X^2+Y^2}{4\sigma^2} \right)}I_{0}\left(\frac{XY}{2\sigma^2} \right) \right]
\end{aligned}
\label{eq:noncoherent30_noncohrent_ber}
\end{equation}
where again $Y=AT\sqrt{{\rho_{10}}_\mathfrak{R}^2+{\rho_{10}}_\mathfrak{I}^2}$ and $X=AT$. This can be expressed in terms of received signal energy by replacing A with $\sqrt{\frac{2E_s}{T}}$.\par
As in the coherent case, by expanding the derivation to include $N$ users in the system, it is relatively straightforward to find a general BER expression for the N-user noncoherent CSS system between any user $k$ and $m$ provided as,
\begin{equation}
\begin{aligned}
P_{b}=\frac{1}{2^{N-1}} 
\left [\begin{multlined}\frac{1}{2}\exp{\left(-\frac{E_{s}\sqrt{1+\sum_{\substack{k=0\\k\neq m}}^{N-1}(\rho_{km_{\Re}}^2+\rho_{km_{\Im}}^2+2\rho_{km_{\Re}}       )}}{2N_{0}} \right)}+\\
\sum_{\xi=1}^{2^{N-1}-1} \left[\begin{multlined}Q\left( \frac{Y_{\xi}\{\pmb{b}_{\xi}\}}{\sigma\sqrt{2}},\frac{X_{\xi}\{\overline{\pmb{b}_{\xi}}\}}{\sigma\sqrt{2}}\right) -\exp{\left(-\frac{X_{\xi}\{\overline{\pmb{b}_{\xi}}\}^2+Y_{\xi}\{{\pmb{b}_{\xi}}\}^2}{4\sigma^4} \right)}\\
\times I_{0}\left(\frac{X_{\xi}\{\overline{\pmb{b}_{\xi}}\}Y_{\xi}\{{\pmb{b}_{\xi}}\}}{2\sigma^2} \right)\end{multlined}\right]\end{multlined}\right]
\end{aligned}
\label{eq:noncoherent31}
\end{equation}
where,
\begin{equation}
\begin{aligned}
&X_{\xi}\{\overline{\pmb{b}_{\xi}}\}=AT\left(\begin{multlined}
1+2(\overline{\pmb{b}_{\xi}^{T}}\times\pmb{\rho_m})^2+2(\overline{\pmb{b}_{\xi}^{T}}\times\pmb{\rho_m})
\end{multlined} \right)\\
&\overline{\pmb{b}_{\xi}}=1-{\pmb{b}_{\xi}}\\
&Y_{\xi}\{{\pmb{b}_{\xi}}\}=AT\sqrt{2(\pmb{b}_{\xi}^{T}\times\pmb{\rho_{m}})^2 }
\end{aligned}
\label{eq:noncoherent32}
\end{equation}
and $\pmb{b}_{\xi}$ is a vector of size $(N-1)\times1$, with superscript $T$ denoting transpose:
\begin{equation}
    \pmb{b}_{\xi}=\left[\begin{matrix}a\left(\xi,0\right)\ \\a\left(\xi,1\right)\\\vdots\\a\left(\xi,N-1\right)\\\end{matrix}\right]
    \label{eq:noncoherent33}
\end{equation}
with $a\left(\xi,i\right)\in\left\{0,1\right\}$ the $i$th coefficient in the binary expansion for decimal number $\xi$, introduced in \eqref{eq:noncoherent34}, and $\pmb{\rho}_{m}$ is the cross correlation vector of dimension $(N-1)\times1$, which is a column of the complete correlation matrix introduced in \eqref{eq:noncoherent36} by changing user index $m$ to $k$.
\par
\begin{figure}[h]

  \includegraphics[width=\linewidth]{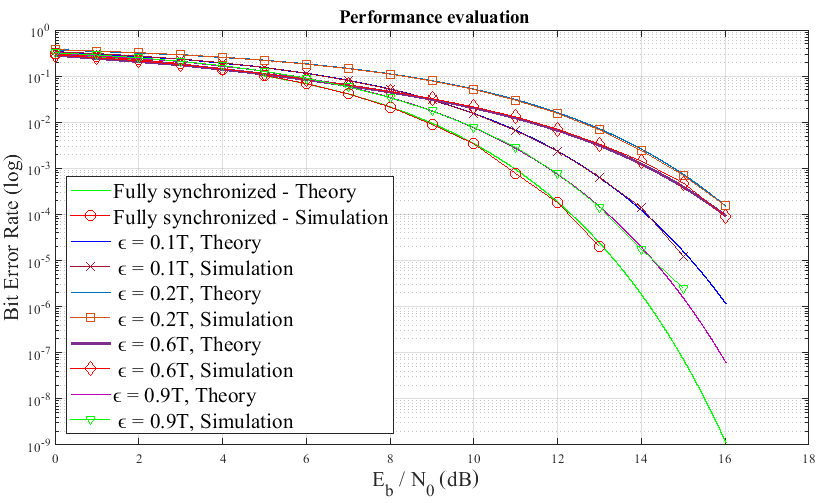}
  \caption{BER vs. SNR for two-user noncoherent linear CSS.}\label{fig:two_user_noncoherent_proof}
  \end{figure}
\begin{figure}[h]
  \includegraphics[width=\linewidth]{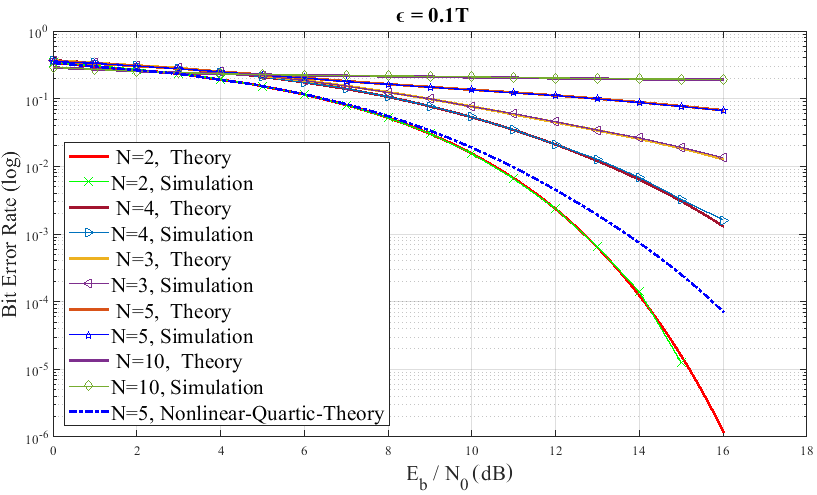}
  \caption{BER vs. SNR for N-user noncoherent linear CSS.}\label{fig:N_user_noncoherent_proof}
\end{figure}

For example for $N$=4 (four user system), $b_{0}=\left[  \begin{matrix}0\\0\\0 \end{matrix}\right]$, $b_{1}=\left[  \begin{matrix}1\\0\\0 \end{matrix}\right]$, $b_{2}=\left[  \begin{matrix}0\\1\\0 \end{matrix}\right]$, $b_{3}=\left[  \begin{matrix}1\\1\\0 \end{matrix}\right]$, $b_{4}=\left[  \begin{matrix}0\\0\\1 \end{matrix}\right]$, $b_{5}=\left[  \begin{matrix}1\\0\\1 \end{matrix}\right]$, $b_{6}=\left[  \begin{matrix}0\\1\\1 \end{matrix}\right]$, $b_{7}=\left[  \begin{matrix}1\\1\\1 \end{matrix}\right]$. As in the coherent case, although our derivation focused on the linear chirp signals, the BER expression is general, and can be used for $any$ specific chirp signaling set, as long as we can compute the cross correlations.
Here, we use the expression for NC BER in Fig. \ref{fig:two_user_noncoherent_proof} and \ref{fig:N_user_noncoherent_proof}. As shown in Fig. \ref{fig:two_user_noncoherent_proof} and \ref{fig:N_user_noncoherent_proof}, analytical results match perfectly with Monte Carlo simulations using MATLAB. Fig. \ref{fig:two_user_noncoherent_proof} presents BER performance for different values of delay (cross correlation values) for a two-user system, and Fig. \ref{fig:N_user_noncoherent_proof} for a fixed delay and different number of users ($N$).

\section{Performance in the Presence of Doppler}\label{sec:Doppler Effect}
\begin{figure}[ht]
	\centering
	{\includegraphics[width =\linewidth]{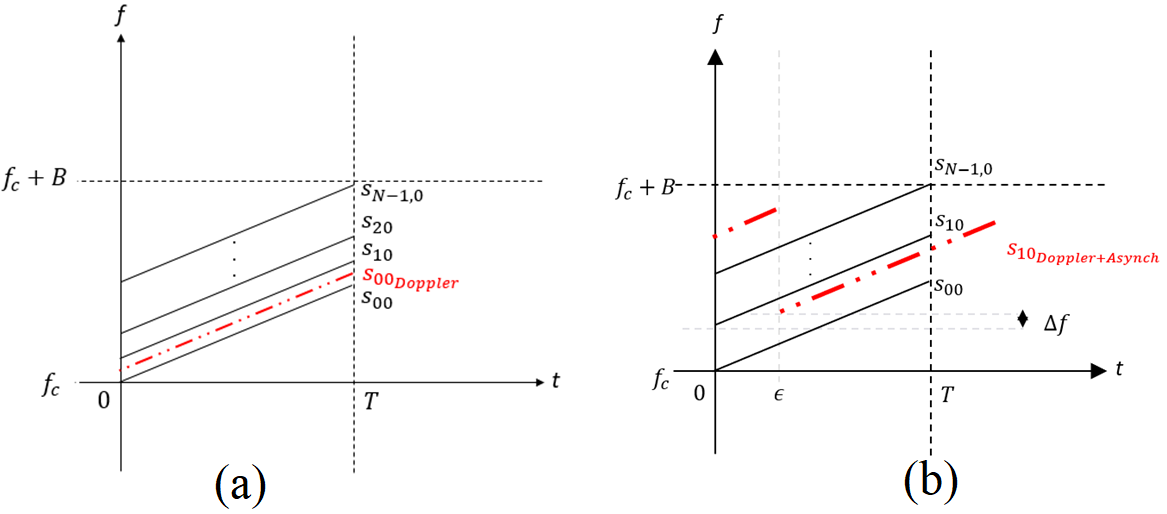}
	}
	\vspace{-3mm}
	\caption{Time-frequency domain representation of (a) Doppler shift, and (b) Doppler and asynchronism effects on a specific linear chirp signal ($s_{10}$). }
	\label{fig:doppler_time_frequency_avali}
\end{figure}

In this section, we investigate the Doppler effect due to transmitter and/or receiver movements. For example, in aeronautical communication the aircraft can travel at high speed, so we expect large Doppler shifts. In our time-frequency representation, we can show the Doppler effect on example linear chirp signals as illustrated in Fig. \ref{fig:doppler_time_frequency_avali} (a). Doppler shift effects can be similar to those of asynchronism, as depicted in Fig. \ref{fig:doppler_time_frequency_avali} (b). %Any given user’s signal experiences a shift in frequency, and if large enough, this shift may cause the signal’s frequency-domain content to overlap with that of other user signals in the system. 

If we consider a Doppler shift of $\Delta f$ for one user ($m$) signal in an $N$ user system, the cross-correlation expression can be written as, 
\begin{equation}
\begin{aligned}
    \rho_{mk}(\Delta f)&= \int_{0}^{T}{exp\left(j\pi\left(\frac{N}{T^2}\left(t+\frac{mT}{N}\right)^2+2\Delta ft\right)\right)}
    %\\&
    \times exp\left( -\frac{j \pi N}{T^2}\left(t+\frac{kT}{N}\right)^2\right) \\
    &=\frac{jT}{2\pi(\Delta f T - k +m)}\times%\\&
    \left(\begin{multlined}[b]\begin{array}{l}exp\left(-j\pi\left(  \frac{k^2-m^2}{N}-2(\Delta f T - k + m)\right)\right)\\
    -exp\left( -j\pi\left(\frac{k^2-m^2}{N}\right)\right) \end{array}\end{multlined}\right).
    \end{aligned}
    \label{eq:Doppler only}
\end{equation}

\begin{figure*}
\centering
\subfloat[]{\includegraphics[width=3.25in]{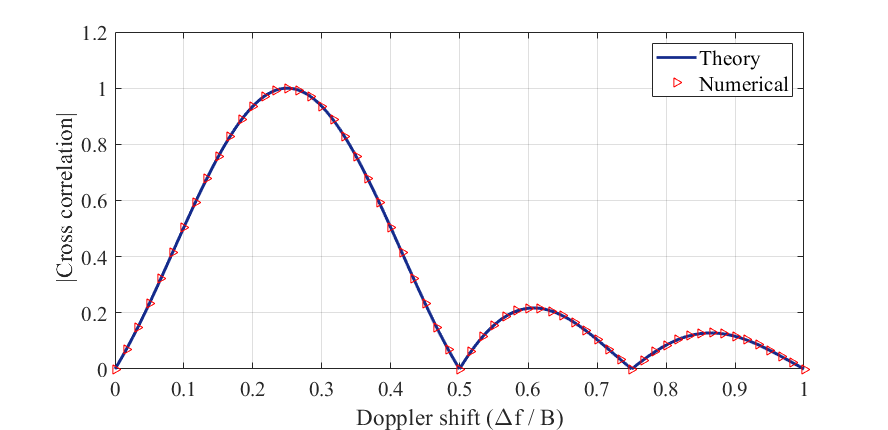}} 
\subfloat[]{\includegraphics[width=3.25in]{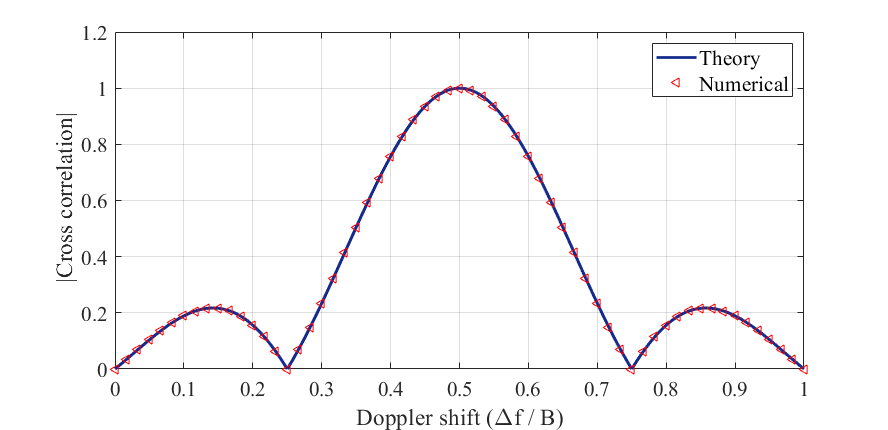}}
\hfill
\subfloat[]{\includegraphics[width=3.25in]{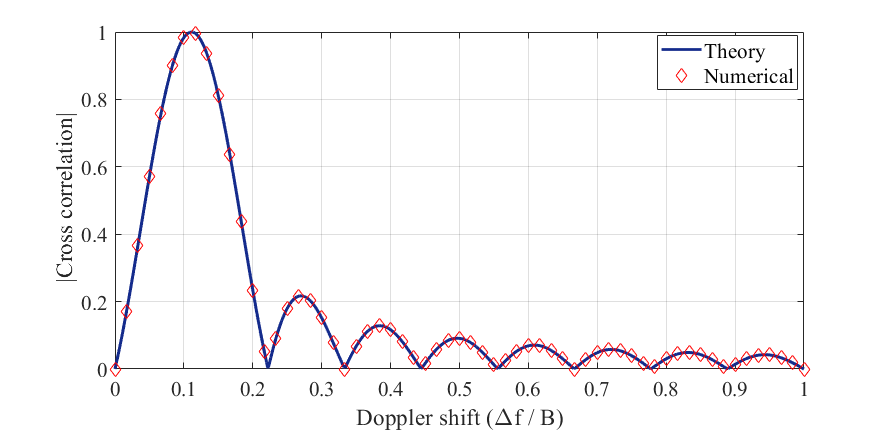}}
\subfloat[]{\includegraphics[width=3.25in]{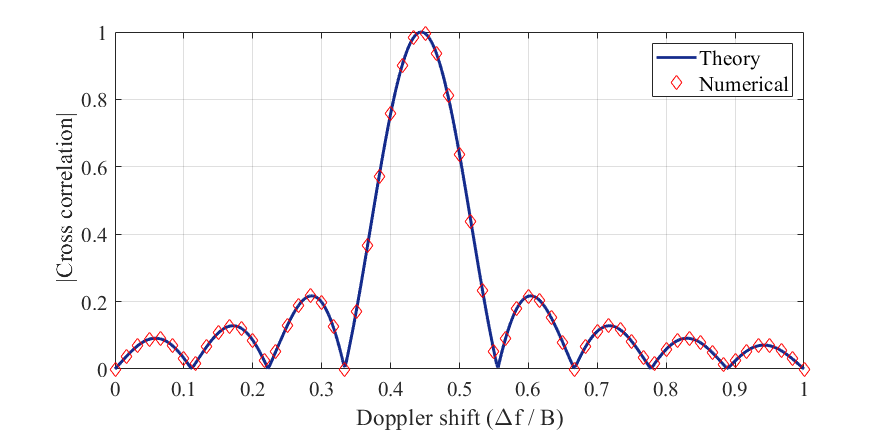}}
\caption{Cross-correlation magnitude vs. normalized Doppler shift (a) user $k=1$ signal experiences due to Doppler shift of user $m=0$ signal, in a five-user system (b) user $k=2$ signal experience due to the Doppler shift of user $m=0$ signal, in a five user system (c) user $k=1$ signal experiences due to the user $m=0$ signal Doppler shift in an $N=10$ user system (d) user $k=5$ signal experiences due to the user $m=0$ signal Doppler shift in an $N=10$ user system.} 
\label{fig:Doppler_only_figures} 
\end{figure*}

To help interpret \eqref{eq:Doppler only}, we consider an example of five and ten users in a set. For the situation in which only the user $m=0$ signal has a Doppler shift, the cross correlation that users k=1, 2 in a set of $N=5$ users, and users $k$ equal to 1 and 5 in a set of $N=10$ users experience for a range of Doppler shifts is plotted in Fig. \ref{fig:Doppler_only_figures} (a)-(d) respectively. We can observe that numerical result match analytical results. 
We can see in Fig. \ref{fig:Doppler_only_figures} (a), as the user $m=0$ signal is Doppler shifted in frequency, the correlation increases to a maximum when the two user signals overlap each other $(\rho_{10}=1)$, and generally when the shift is $(k-m)/N$. As the signal Doppler frequency further increases, this overlap eventually occurs with other user signals, and the cross-correlation with the user 1 signal reaches zero. Note that we normalize the Doppler shift since larger values of Doppler shift have no effect on the signals when they are outside their subband. Similarly, the cross-correlation that user $k=2$ experiences is shown in Fig. \ref{fig:Doppler_only_figures} (b) As expected, the peak occurs when the user 0 signal’s Doppler shift is exactly equal to the frequency difference between user signals 0 and 2.
The analytical and numerical results for $N=10$ user signals, with cross-correlations that user $k=1$ and $k=5$ signals experience, are plotted in Fig. \ref{fig:Doppler_only_figures} (c) and (d), respectively. As with the previous two figures, we can predict the peak values at Doppler shift values corresponding to complete overlap in frequency.
\begin{figure}[ht!]
	\centering
	{\includegraphics[width =\linewidth]{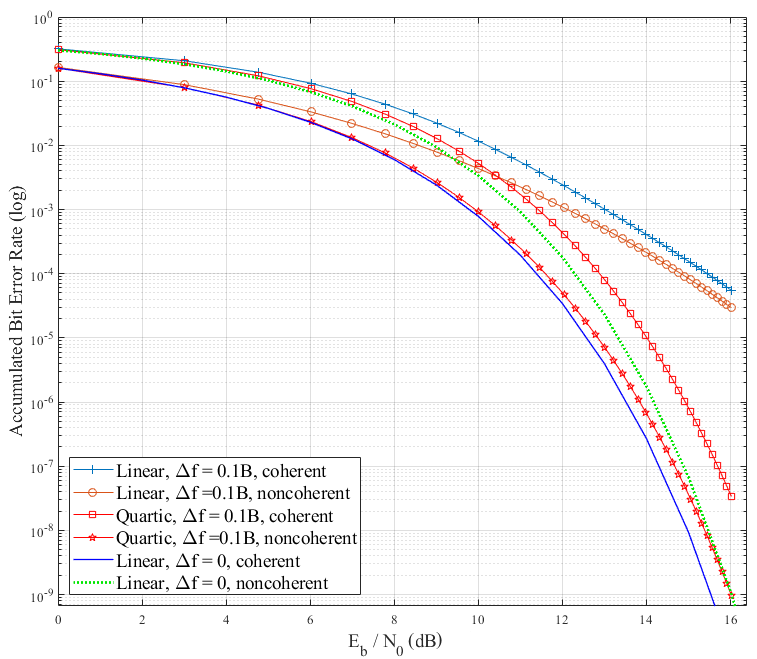}
	}
	
	\caption{BER vs. $E_b/N_0$ for the two chirp waveforms (linear, and quartic) for different values of Doppler shift that user $k=3$ experiences for $N=12$ when the $N-1$ other users are Doppler shifted.}
	\label{fig:doppler_BER}
\end{figure}
Two nonlinear chirp candidates (sinusoidal and quartic) introduced in \cite{journal1}, were shown to have better performance in the coherent case. We can see in Fig. \ref{fig:doppler_BER} how the quartic signal case improves performance for noncoherent and coherent detection in the presence of Doppler.
We can also observe in Fig. \ref{fig:doppler_BER} that noncoherent BER performance under Doppler shifts for both linear and quartic case is better than coherent case. This interesting finding is another topic for future investigations.
\section{Combined Doppler and Asynchronism  Effects}\label{sec:Doppler and Asynchronous Effect}
In this section, we investigate the combined effects of Doppler and asynchronism on performance of BCSS. In our time-frequency representation, we can show this combined effect on example linear chirp signals as illustrated in Fig. \ref{fig:doppler_time_frequency_avali} (b). Any given user’s signal experiences a shift in frequency and delay in time, and for specific values, this shift or delay may cause the signal’s frequency-domain content to overlap with that of other user signals in the system. 
If we consider a Doppler shift of $\Delta f$  and delay of $\epsilon$ for one user ($m$) signal in an $N$ user system, the cross-correlation expression can be written as follows,

\begin{equation}
\begin{aligned}
    &\rho_{mk}(\Delta f, \epsilon)= \int_{0}^{T}{exp\left(j\pi\left(\frac{N}{T^2}\left(t+\frac{mT}{N}-\epsilon\right)^2+2\Delta ft\right)\right)}
    \\
    &\times exp\left( -\frac{j \pi N}{T^2}\left(t+\frac{kT}{N}\right)^2\right) .
    \end{aligned}
    \label{eq:noncoherent37}
\end{equation}

As previously described in \cite{journal1}, we have to divide the integral into two parts and the final cross correlation expression can be found in (\ref{eq:noncoherent38}).

\begin{figure*}[h!]
\begin{equation}
\begin{aligned}
     &\rho_{mk}(\Delta f, \epsilon)=\frac{jT}{2 \pi (\Delta f T^2-kT+mT-N\epsilon)}\\
     &\times\left[\begin{multlined}[b] \begin{array}{l} exp\left(-j\pi\left(\frac{k^2T^2+2kNT^2-m^2T^2+2mNT(\epsilon-T)+N(N\epsilon(2T-\epsilon)-2\Delta T^3)}{NT^2} \right) \right)\\
     -exp\left(-j\pi\left( \frac{k^2T^2+2kNT\epsilon-m^2T^2+N(N\epsilon^2+2\Delta fT^2 \epsilon}{NT^2}\right) \right)\end{array}\end{multlined}\right]\\
     &+\frac{jT}{2 \pi \left(\Delta f T^2-kT+mT-N(-T+\epsilon)\right)}\\
     &\times\left[\begin{multlined}[b] \begin{array}{l} exp\left(-j\pi\left(\frac{k^2T^2+2kNT\epsilon-m^2T^2+2mNT(-T)+N(N(-T+\epsilon)(\epsilon)-2\Delta fT^2\epsilon)}{NT^2} \right) \right)\\
     -exp\left(-j\pi\left( \frac{k^2T^2-m^2T^2+2mNT(-T+\epsilon)+N(N(-T+\epsilon)(\epsilon))}{NT^2}\right) \right)\end{array}\end{multlined}\right]
     \end{aligned}
    \label{eq:noncoherent38}
\end{equation}
\end{figure*}

\begin{figure*}[ht!]
\centering
\subfloat[]{\includegraphics[width=3.25in]{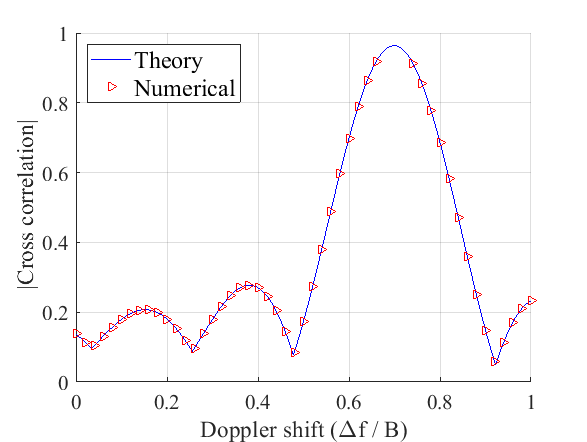}} 
\subfloat[]{\includegraphics[width=3.25in]{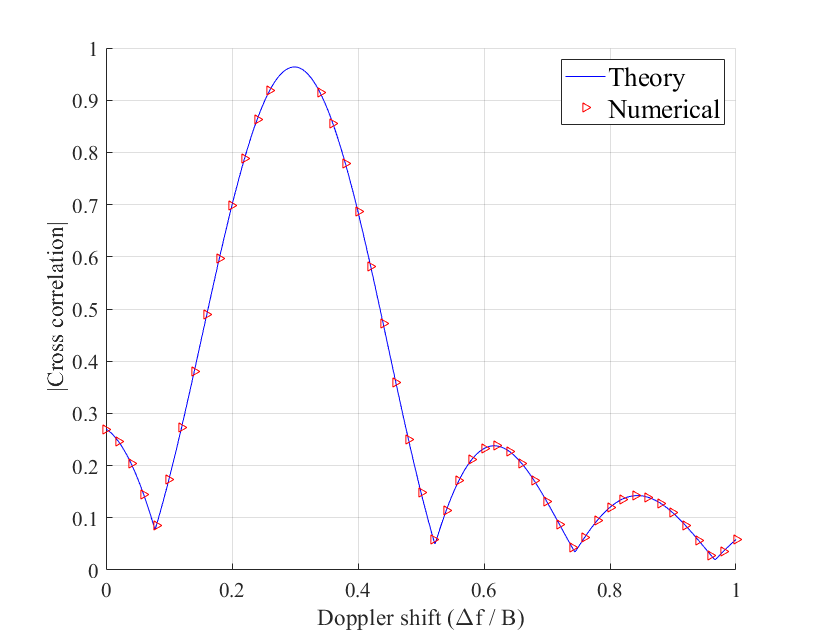}}
\hfill
\subfloat[]{\includegraphics[width=3.25in]{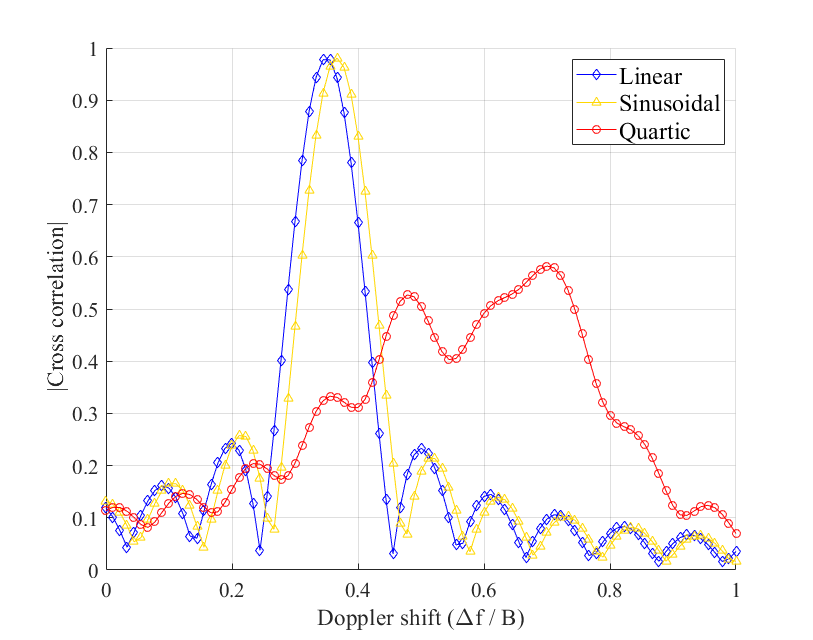}}
\subfloat[]{\includegraphics[width=3.25in]{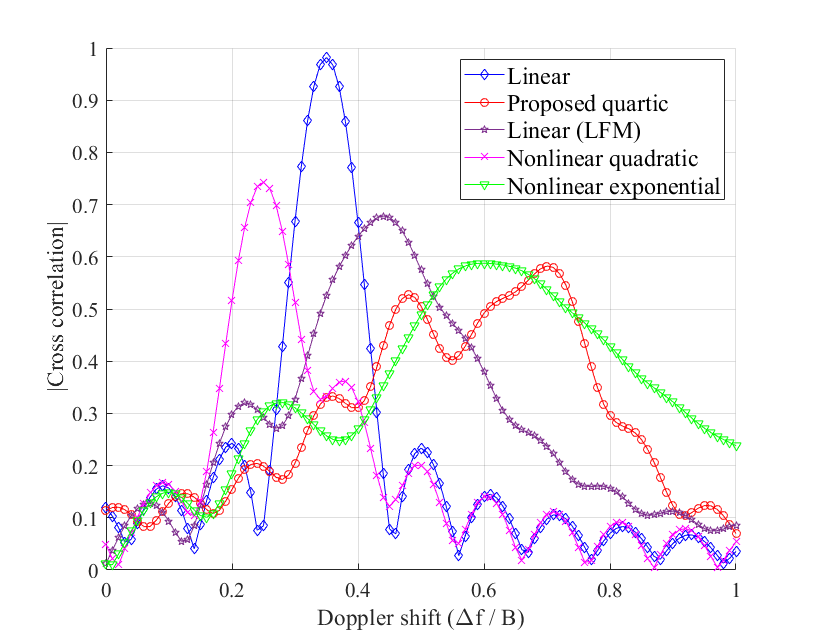}}
\caption{Cross-correlation magnitude vs. normalized Doppler shift (a) for user $k=4$ signal experiences due to Doppler shift of user $m=0$ signal, in a five-user system (b) user $k=2$ signal experience due to the Doppler shift plus asynchronism of the user $m=0$ signal, for $\epsilon=0.1T$ in an $N=5$ user system (c) comparison of user $k=4$ due to the Doppler shift plus asynchronism of proposed nonlinear chirps in \cite{journal1} for $\epsilon=0.05T$ and $N=10$ (d) comparison of user $k=4$ due to the Doppler shift plus asynchronism of several other chirp designs in the literature, i.e., linear LFM \cite{LINEAR_LFM}, nonlinear quadratic and nonlinear exponential \cite{KHAN_NONLINEAR_QUADRATIC}.} 
\label{fig:EcUND} 
\end{figure*}

Figure \ref{fig:EcUND} shows cross-correlation that (a) user two and (b) user four experiences from the Doppler shift of the user $m=0$ signal for $\epsilon=0.1T$. An interesting observation is that we no longer see perfectly orthogonal $(\rho_{mk}=0)$ conditions for any Doppler shifts because of the time-shifting. Similarly we do not experience full overlap $(\rho=1)$. This leads to a hypothesis that Doppler shift along with asynchronism may prevent peak cross-correlations. Analogous results presented for a system with larger values of $N$ appear in \cite{hosseini2020novel}. 
\par
We investigate our proposed nonlinear chirp designs in these conditions, and compare with the linear and other nonlinear chirps in  the literature, so we selected linear and nonlinear quadratic and exponential chirps as introduced in \cite{LINEAR_LFM} and \cite{KHAN_NONLINEAR_QUADRATIC} respectively. The cross correlations versus normalized Doppler shift for $\epsilon=0.1T$ and $N=5$ are plotted in Fig. \ref{fig:EcUND}. In these graphs, we plot the correlation that the user $k=4$ signal experiences when user signal $m=0$ has been both Doppler shifted and is asynchronous. As we can see in Fig. \ref{fig:EcUND} (c), the nonlinear quartic case has lower peaks but a broader main lobe in its values of cross-correlation. In Fig. \ref{fig:EcUND} (d), we illustrate the correlations for linear and some nonlinear chirps from the literature to show that the nonlinear chirps have this reduced-correlation-peak advantage over the typical linear chirp. Note that as mentioned in \cite{journal1}, the other nonlinear chirps also have other disadvantages in comparison to our proposed nonlinear signals, such as amplitude variation (larger peak-to-average power ratio), and significant bandwidth variation among different user signals. 
\par
\begin{figure}[ht!]
	\centering
	{\includegraphics[width =\linewidth]{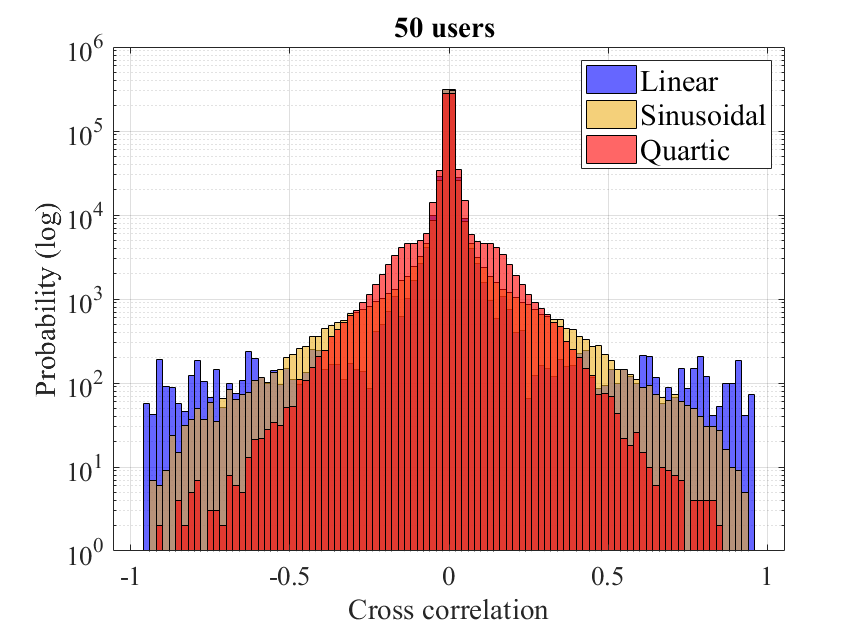}
	}
	
	\caption{Cross-correlation that all users experience due to the Doppler shift plus asynchronism of $\epsilon=0.05T$, for $N=50$ user signals.}
	\label{fig:histogram}
\end{figure}

Figure. \ref{fig:EcUND} (c) shows quartic and sinusoidal chirp cross correlations have smaller values than linear, but for specific delay values we may experience worse BER due to higher value of cross correlation. Therefore, we plotted a histogram of all cross correlation values that all users in a set of $N=50$ users and delay of $\epsilon=0.05T$ in Fig. \ref{fig:histogram}. We can see that the quartic case has smaller maximum cross correlation values overall.  The correlation standard deviations for N=50 are 0.0714, 0.0691 and 0.0614 for linear, sinusoidal and quartic respectively.

\section{Conclusion}\label{sec:conclusion}
In this paper, multi-user chirp spread spectrum non-coherent system performance with both asynchronism and Doppler shift conditions has been investigated, both for the classic linear and two new nonlinear chirps. We previously derived a closed form expression for the cross correlation and coherent bit error ratio for linear and nonlinear chirps in \cite{journal1}. These BER expressions are general, and can be used for arbitrary chirp signal designs as long as cross-correlations can be computed.
We validated our analysis via numerical and simulation results, and provided cross correlation derivation for both Doppler and asynchronism. The linear chirps are generally best in perfectly synchronized cases, but we showed in \cite{journal1} that since our nonlinear cases use more “time-frequency space,” they can outperform all linear and nonlinear chirp designs we have evaluated, for a range of assumed Doppler or delay offsets.  The BER performance of our proposed quartic chirps is superior to that of the linear chirps for these investigated conditions.  Finally, to evaluate a more realistic situation, we imposed both Doppler and asynchronism together, and evaluated our coherent and noncoherent receiver performance.

\section{Acknowledgment}\label{sec:acknowledgment}
The authors would like to thank the reviewers for their constructive comments. Views expressed herein are those of the authors and may not represent the views of the sponsoring agency. The work in this paper was led by D. W. Matolak. N. Hosseini drafted the paper and performed all the computer simulations. Both authors were involved with all analyses and in revising and editing the paper. All authors declare no competing interests.

\appendix \setcounter{equation}{0} \renewcommand\theequation{A.\arabic{equation}}Here we find a solution for the following integral,
  \begin{equation}
      I=\int_{0}^{\infty}{x\exp{\left(-\frac{p^2x^2}{2}\right)}I_0\left(cx\right)Q_1\left(\beta,\alpha x\right)dx}
            	\label{eq.A1}
  \end{equation}
  We know that,
  \begin{equation}
      Q\left(A,B\right)+Q\left(B,A\right)=1+\exp{\left(-\frac{A^2+B^2}{2}\right)}I_0\left(AB\right).
      	\label{eq.A2}
  \end{equation}
  By substituting \eqref{eq.A2} into \eqref{eq.A1},
  \begin{equation}
  \begin{aligned}
      I&=\int_{0}^{\infty}{x\exp{\left(-\frac{p^2x^2}{2}\right)}I_0\left(cx\right)}\\
     &\times \left(1+\exp{\left(-\frac{{\alpha^2x}^2+\beta^2}{2}\right)
      I_0\left(axb\right)-Q\left(\alpha x,\beta\right)}\right)dx
      \\
      &=\int_{0}^{\infty}{x\exp{\left(-\frac{p^2x^2}{2}\right)}I_0\left(cx\right)}\\
      &+\int_{0}^{\infty}{x\exp{\left(\frac{-\left(\alpha^2+p^2\right)x^2+\beta^2}{2}\right)}I_0\left(cx\right)}I_0\left(axb\right)\\
      &-\int_{0}^{\infty}{x\exp{\left(-\frac{p^2x^2}{2}\right)}I_0\left(cx\right)}Q\left(\alpha x,\beta\right)dx.
      \end{aligned}
  	\label{eq.A3}
  \end{equation}
  From the definition of the Marcum Q-function we can obtain,
  \begin{equation}
      \int_{b}^{\infty}{x\exp{\left(-\frac{p^2x^2}{2}\right)I_0\left(Ax\right)=\frac{1}{p^2}\exp{\left(\frac{A^2}{2p^2}\right)}}}Q\left(\frac{A}{p},bp\right).
      \label{eq.A4}
  \end{equation}
  Therefore by using \eqref{eq.A4} in \eqref{eq.A3}, 
  \begin{equation}
  \begin{aligned}
      I&=\frac{1}{p^2}\exp{\left(\frac{c^2}{2p^2}\right)}Q\left(\frac{c}{p},0\right)\\
      &+\int_{0}^{\infty}{x\exp{\left(-\frac{\left(\alpha^2+p^2\right)x^2+\beta^2}{2}\right)}I_0\left(cx\right)}I_0\left(axb\right)\\
      &-\int_{0}^{\infty}{x\exp{\left(-\frac{p^2x^2}{2}\right)}I_0\left(cx\right)}Q\left(\alpha x,\beta\right)dx\ .
         \label{eq.A5}
      \end{aligned}
  \end{equation}
  We know $Q\left(a,0\right)=1$, therefore with some mathematical simplification,
    \begin{equation}
  \begin{aligned}
     I&=\frac{1}{p^2}\exp{\left(\frac{c^2}{2p^2}\right)}\\
     &+\exp{\left(\frac{\beta^2}{2}\right)}\int_{0}^{\infty}{x\exp{\left(-\frac{\left(\alpha^2+p^2\right)x^2}{2}\right)}I_0\left(cx\right)}I_0\left(ax\beta\right)\\
     &-\int_{0}^{\infty}{x\exp{\left(-\frac{p^2x^2}{2}\right)}I_0\left(cx\right)}Q\left(\alpha x,\beta\right)dx\ .
         \label{eq.A6}
      \end{aligned}
  \end{equation}
  For the first integral with two modified Bessel functions, we know from \cite{appendix_ref}, 
      \begin{equation}
  \begin{aligned}
\int_{0}^{\infty}&xe^{\left(-\tau^2x^2\right)}I_0\left(Ax\right)I_0\left(Bx\right)dx\\
&={{\frac{1}{2\tau^2}\exp{\left(-\frac{A^2+B^2}{4\tau^2}\right)}}I_0\left(\frac{AB}{2\tau^2}\right)}\\
&for\ A>0,\ B>0\ and\ \left|\arg{\ \tau}\right|<\frac{\pi}{4}\ .
\label{eq.A7}
      \end{aligned}
  \end{equation}
  By using \eqref{eq.A7} in the second term of \eqref{eq.A6} we will have,
       \begin{equation}
  \begin{aligned}
I&=\frac{1}{p^2}\exp{\left(\frac{c^2}{2p^2}\right)}\\
&+\exp{\left(\frac{\beta^2}{2}\right)}\frac{1}{\alpha^2+p^2}\exp{\left(-\frac{a^2\beta^2+c^2}{2\left(\alpha^2+p^2\right)}\right)I_0\left(\frac{a\beta c}{\alpha^2+p^2}\right)}\\
&-\int_{0}^{\infty}{x\exp{\left(-\frac{p^2x^2}{2}\right)}I_0\left(cx\right)}Q\left(\alpha x,\beta\right)dx.\ 
\label{eq.A8}
      \end{aligned}
  \end{equation}
  Now we denote the third term of \eqref{eq.A8} $F$, 
  \begin{equation}
      F=\int_{0}^{\infty}{x\exp{\left(-\frac{p^2x^2}{2}\right)}I_0\left(cx\right)}Q\left(\alpha x,\beta\right)dx.
  \end{equation}
  From Marcum Q function properties we have, \begin{equation}
      \frac{\partial Q\left(A,B\right)}{\partial\beta}=-B\exp{\left(-\frac{A^2+B^2}{2}\right)}I_0\left(AB\right),\ 
  \end{equation}
  then by taking the derivative of $F$ with respect to $\beta$, we can write, 
  \begin{equation}
      \begin{aligned}
      \frac{\partial F}{\partial\beta}&=\int_{0}^{\infty}{x\exp{\left(-\frac{p^2x^2}{2}\right)}I_0\left(cx\right)}\\
      &\times\left(-\beta\exp{\left(-\frac{{\alpha^2x}^2+\beta^2}{2}\right)I_0\left(\alpha x\beta\right)}\right)\\
&=\int_{0}^{\infty}{x\exp{\left(-\frac{(\alpha^2+p^2)x^2}{2}\right)}I_0\left(cx\right)}\\
&\times\left(-\beta\exp{\left(-\frac{\beta^2}{2}\right)I_0\left(\alpha x\beta\right)}\right)\ \ 
\label{eq.A11}
      \end{aligned}
  \end{equation}
  Again by using \eqref{eq.A7} we will have,
  \begin{equation}
  \begin{aligned}
      \frac{\partial F}{\partial\beta}=\frac{-\beta}{\alpha^2+p^2}\exp{\left(-\frac{a^2\beta^2+c^2}{2\left(\alpha^2+p^2\right)}\right)}I_0\left(\frac{\alpha\beta c}{\alpha^2+p^2}\right)\\
      \times\left(\exp{\left(-\frac{\beta^2}{2}\right)}\right)\ \ .
      \label{eq.A12}
      \end{aligned}
  \end{equation}
  Since $F$=0 when $\beta=\infty$, simplifying \eqref{eq.A12} yields, 
  \begin{equation}
      \begin{aligned}
F&=\int_{\beta}^{\infty}{\frac{-x}{\alpha^2+p^2}\exp{\left(\frac{-p^2x^2+c^2}{2\left(\alpha^2+p^2\right)}\right)}I_0\left(\frac{\alpha xc}{\alpha^2+p^2}\right)dx}\\
&=\frac{-1}{\alpha^2+p^2}\exp{\left(\frac{c^2}{2\left(\alpha^2+p^2\right)}\right)}\\
&\times\int_{\beta}^{\infty}{x\exp{\left(-\frac{p^2x^2}{2\left(\alpha^2+p^2\right)}\right)}}
 {I_0\left(\frac{\alpha xc}{\alpha^2+p^2}\right)dx}\ .
\label{eq.A13}
\end{aligned}
  \end{equation}
  By using \eqref{eq.A4} again we obtain,
  \begin{equation}
      \begin{aligned}
      F&=\frac{-1}{\alpha^2+p^2}\exp{\left(\frac{c^2}{2\left(\alpha^2+p^2\right)}\right)}\frac{\left(\alpha^2+p^2\right)}{p^2}\\
      &\times \exp{\left(\frac{a^2c^2}{2p^2(\alpha^2+p^2)}\right)} Q\left(\frac{ac}{p\sqrt{\left(\alpha^2+p^2\right)}},\frac{\beta p}{\sqrt{\left(\alpha^2+p^2\right)}}\right)\\
      &=\frac{-1}{p^2}\exp{\left(\frac{c^2}{2p^2}\right)}\ Q\left(\frac{ac}{p\sqrt{\left(\alpha^2+p^2\right)}},\frac{\beta p}{\sqrt{\left(\alpha^2+p^2\right)}}\right).
      \label{eq.A14}
      \end{aligned}
  \end{equation}
  Returning to our derivation by substituting \eqref{eq.A14} into \eqref{eq.A8} we will have, 
   \begin{equation}
      \begin{aligned}
I&=\frac{1}{p^2}\exp{\left(\frac{c^2}{2p^2}\right)} \\
&-\frac{1}{p^2}\exp{\left(\frac{c^2}{2p^2}\right)}\ Q\left(\frac{ac}{p\sqrt{\left(\alpha^2+p^2\right)}},\frac{\beta p}{\sqrt{\left(\alpha^2+p^2\right)}}\right)\\
&+\exp{\left(-\frac{\beta^2}{2}\right)}\frac{1}{\alpha^2+p^2}\exp{\left(-\frac{a^2\beta^2+c^2}{2\left(\alpha^2+p^2\right)}\right)}\\
&{\times I_0\left(\frac{a\beta c}{\alpha^2+p^2}\right)}\\
&=\frac{1}{p^2}\exp{\left(\frac{c^2}{2p^2}\right)}\\
&-\frac{1}{p^2}\exp{\left(\frac{c^2}{2p^2}\right)}\ Q\left(\frac{ac}{p\sqrt{\left(\alpha^2+p^2\right)}},\frac{\beta p}{\sqrt{\left(\alpha^2+p^2\right)}}\right)\\
&+\frac{1}{\alpha^2+p^2}\exp{\left(\frac{-p^2\beta^2+c^2}{2\left(\alpha^2+p^2\right)}\right)I_0\left(\frac{a\beta c}{\alpha^2+p^2}\right)}
      \label{eq.A14}
      \end{aligned}
  \end{equation}
  Then by using \eqref{eq.A2},
  \begin{equation}
      \begin{aligned}
     I&=\left(\frac{-\alpha^2}{p^2\left(\alpha^2+p^2\right)}\right) \exp{\left(\frac{{-p}^2\beta^2+c^2}{2\left(\alpha^2+p^2\right)}\right)I_0\left(\frac{a\beta c}{\alpha^2+p^2}\right)}\\
     &+\frac{1}{p^2}\exp{\left(\frac{c^2}{2p^2}\right)}Q\left(\frac{\beta p}{\sqrt{\left(\alpha^2+p^2\right)}},\frac{ac}{p\sqrt{\left(\alpha^2+p^2\right)}}\right)
      \end{aligned}
  \end{equation}
  Finally, we can write, 
   \begin{equation}
      \begin{aligned}
I=\frac{1}{p^2}\left[\begin{multlined}\exp{\left(\frac{c^2}{2p^2}\right)}Q\left(\frac{\beta p}{\sqrt{\left(\alpha^2+p^2\right)}},\frac{ac}{p\sqrt{\left(\alpha^2+p^2\right)}}\right)\\
-\left(\frac{\alpha^2}{\left(\alpha^2+p^2\right)}\right)\exp{\left(\frac{{-p}^2\beta^2+c^2}{2\left(\alpha^2+p^2\right)}\right)}\\
{\times I_0\left(\frac{a\beta c}{\alpha^2+p^2}\right)}\end{multlined}\right]
      \end{aligned}
  \end{equation}

\bibliographystyle{IEEEtran}
\bibliography{bandlimitedChirps}

\end{document}